\documentclass[pdflatex,sn-mathphys-num,twocolumn]{sn-jnl}
\usepackage{graphicx}%
\usepackage{multirow}%
\usepackage{amsmath,amssymb,amsfonts}%
\usepackage{amsthm}%
\usepackage{mathrsfs}%
\usepackage[title]{appendix}%
\usepackage{xcolor}%
\usepackage{textcomp}%
\usepackage{manyfoot}%
\usepackage{booktabs}%
\usepackage{algorithm}%
\usepackage{algorithmicx}%
\usepackage{algpseudocode}%
\usepackage{listings}%
\usepackage{geometry}
\usepackage{lineno}
\begin{document}
\title[Particle species dependence of femtoscopic source parameters]{Particle species dependence of femtoscopic source parameters in high-energy nuclear collisions}

\author*[1]{\fnm{Dániel} \sur{Kincses}}\email{kincses@ttk.elte.hu}
\author[1]{\fnm{László} \sur{Kovács}}
\author[1]{\fnm{Máté} \sur{Csanád}}

\affil[1]{\orgdiv{Department of Atomic Physics}, \orgname{ELTE Eötvös Loránd University}, \orgaddress{\street{Pázmány Péter sétány 1/A}, \city{Budapest}, \postcode{H-1117}, \country{Hungary}}}

\abstract{High-energy nuclear physics explores the properties of strongly interacting matter created in relativistic collisions of nuclei. Femtoscopy, a subfield of high-energy physics, utilizes quantum-statistical correlations of particles to characterize the space-time geometry of the particle-emitting source. Recent measurements and phenomenological investigations indicated that the shape of the source for identical pions can be well-described by Lévy-stable distributions. The significant power-law tail of the pion source observed both in experiment and in simulations has been shown to originate from the process of Lévy walk during the hadronic scattering phase of the collisions. To better understand the physical processes behind the formation of such power laws, an important next step is to investigate particle species dependence, especially the source shape of identical kaon and proton pairs. As a direct continuation of our previous studies, in this paper, we present a detailed three-dimensional investigation of the two-particle source shape in simulations of Au+Au collisions at 200 GeV per nucleon pair collision energy using the EPOS3 model. We show the dependence of the extracted femtoscopic source parameters on particle species, centrality and average transverse mass. We find that the scale parameters do not show a clear transverse mass scaling between particle species, as there are systematic differences in the overlapping regions. The power-law exponents of pion and kaon pairs are compatible, while for protons it is higher, closer to the Gaussian limit. When new experimental measurements of kaon and proton correlations become available, these results will provide the basis of a data-model comparison.}



\maketitle

\section{Introduction}\label{s:intro}

Femtoscopy, in particular two-particle interferometry, has been an essential tool in high-energy physics since the discovery of the quantum-statistical correlations of pion pairs by Goldhaber and collaborators~\cite{Goldhaber:1959mj,Goldhaber:1960sf}. The methodology of femtoscopic correlation measurements, developed over many years of heavy-ion physics research, allows one to explore the space-time geometry of the fireball created in highly energetic collisions of nuclei~\cite{Lisa:2005dd}. One main purpose of such studies is to reconstruct the shape and size of the particle-emitting source, or more specifically, of two-particle spatial correlations. This quantity cannot be directly measured in experiments, but it can be reconstructed by utilizing quantum-statistical effects and the connection to relative-momentum correlations. Recent experimental~\cite{PHENIX:2024vjp,Kincses:2024sin,NA61SHINE:2023qzr,CMS:2023xyd} and phenomenological~\cite{Kincses:2024lnv,Csanad:2024jpy,Nagy:2023zbg,Korodi:2022ohn,Kincses:2019rug,Kincses:2025iaf,Porfy:2025sev,Nagy:2026mvg} studies showed that the shape of the pair source distribution of pions can be described by elliptically-contoured three-dimensional Lévy-stable distributions, exhibiting a prominent power-law tail. In Ref.~\cite{Kincses:2024lnv}, it was shown that these power-law tails originate from the process of Lévy walk, a type of random walk where the appearance of occasional long steps spreads out the spatial distribution to large distances. This happens mainly during the hadronic scattering phase of the collisions, due to intertwined processes such as elastic and inelastic scatterings and resonance formations and decays. In Ref.~\cite{Csanad:2007fr}, it was argued that elastic scattering of hadrons can create anomalous diffusion as well, which could be tested by comparing the source shapes of different particle species, having different scattering cross-sections. In this paper, building on our previous results~\cite{Kincses:2024lnv,Kincses:2025izu,Huang:2025edi}, we utilize state-of-the-art Monte Carlo simulations of Au+Au collisions at a collision energy of 200 GeV per nucleon pair and investigate the shape of the two-particle source of pion, kaon, and proton pairs. Experimental measurements of kaon-kaon and proton-proton correlations utilizing a Lévy-stable source distribution are not yet available (but are currently ongoing at different experiments), so our results provide important input and possible comparison to future data. The structure of this paper is the following. In Section~\ref{s:methods} we discuss the details of the simulation and the applied methods. In Section~\ref{s:results} we present our findings and discuss the particle species, centrality, and average transverse momentum-dependence of the extracted source parameters. In Section~\ref{s:summary} we summarize and conclude.

\section{Methods}\label{s:methods}

The analysis presented here is a direct continuation of Ref.~\cite{Kincses:2025izu}, where only identical charged pions were investigated. We use the exact same dataset, a sample of 300,000 minimum bias Au+Au events simulated at the center-of-mass collision energy of $\sqrt{s_{\rm NN}} = 200\textnormal{ GeV}$ with the EPOS model~\cite{Werner:2010aa} (version 359). Details about the event generator model and the analysis methodology can be found in Ref.~\cite{Kincses:2025izu}; here we focus on the important differences and new developments. The goal of the analysis is to construct the $D(\boldsymbol{\rho})$ spatial separation distributions of identical charged pion, kaon, and proton pairs, where $\boldsymbol{\rho}$ is the three-dimensional spatial separation vector at the point of the last interaction (freeze-out) of the particles. We then compare the respective source parameters of these distributions, obtained by fitting three-dimensional Lévy-stable distributions to them. 


The components of the $\boldsymbol{\rho}$ spatial separation vector are calculated in the Longitudinal Co-Moving System (LCMS) and rotated to the ``out--side--long" coordinates, as defined by Equations~13--15 of Ref.~\cite{Kincses:2024lnv}. The single particle selection criteria were similar to Ref.~\cite{Kincses:2025izu}, i.e., the accepted pseudorapidity range was $|\eta| < 1$, and the momentum range was ${0.15 < p_T~[{\rm GeV}/c]<1.0}$ for pions and kaons, and ${0.15 < p_T~[{\rm GeV}/c]<2.5}$ for protons. The two-particle selection criterion on the magnitude of the relative momentum vector was identical to that of Ref.~\cite{Kincses:2025izu}, i.e., $Q_{\rm LCMS} < \sqrt{(0.15~\textnormal{GeV})\,m_T}$, where $m_T = \sqrt{m^2+k_T^2}$ is the average transverse mass of the pair, $m$ is the particle mass, and $k_T$ is the average transverse momentum of the pair. 
The investigated $k_T$ range for pion pairs was ${0.175 < k_T~[{\rm GeV}/c] < 0.925}$ (15 evenly spaced bins), increasing the upper limit compared to our previous work~\cite{Kincses:2025izu} in order to obtain a wider overlap in transverse mass $m_T$ of pion and kaon source parameters. For kaons and protons the investigated range was ${0.25 < k_T~[{\rm GeV}/c] < 0.95}$ (7 evenly spaced bins) and ${0.7 < k_T~[{\rm GeV}/c] <1.5}$ (8 evenly spaced bins), respectively. For kaon and proton pairs, it was not feasible to further extend the $k_T$ range due to limited statistics. As the number of kaon and proton pairs is much less than that of pions, we limited our range of investigation in centrality to the three most-central classes of $0{-}10\%$, $10{-}20\%$, and $20{-}30\%$.

\begin{figure}
    \centering
    \includegraphics[trim={0 5 35 38}, clip, width=\linewidth]{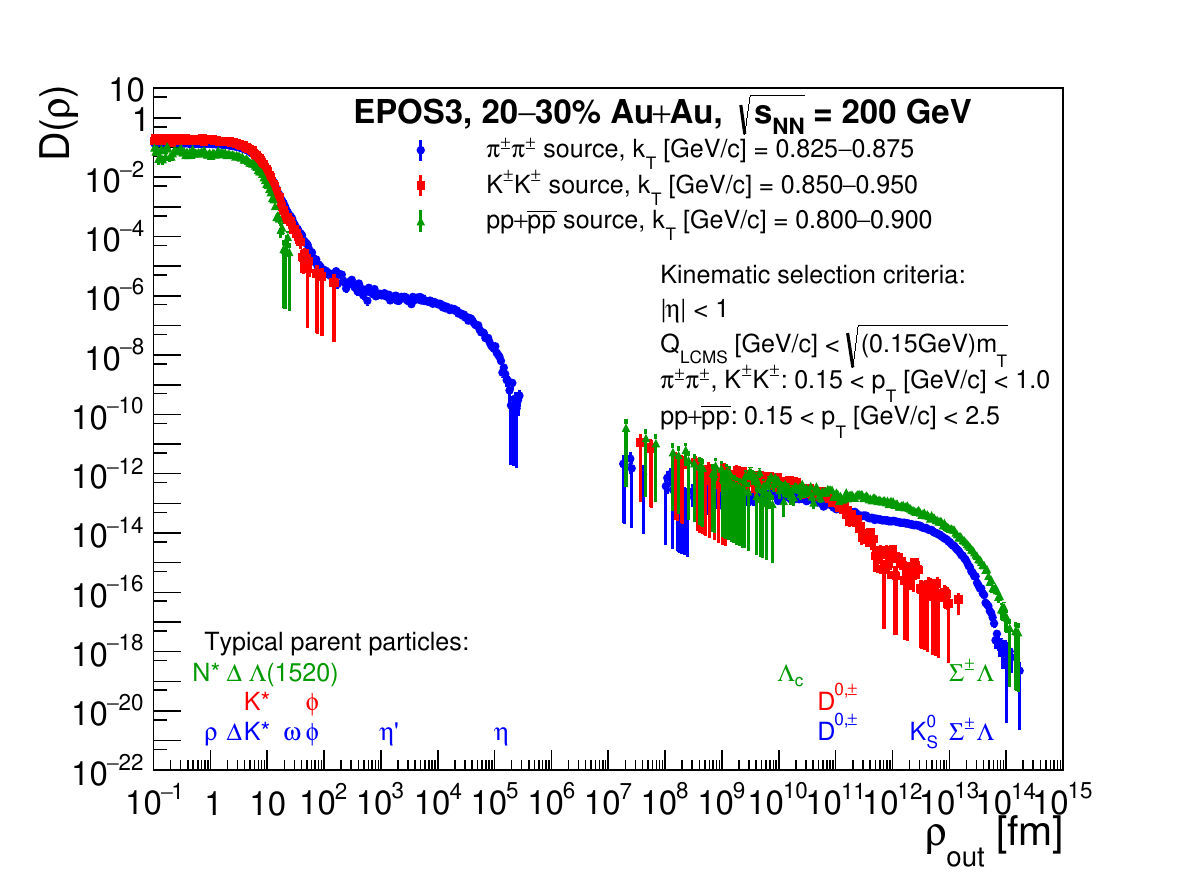}
    \caption{Example $D(\boldsymbol{\rho})$ distributions of pion, kaon, and proton pairs, in the $20{-}30\%$ centrality class, projected in the \textit{out} direction, shown with blue, red, and green markers, respectively. Above the horizontal axes typical parent particles in the given $\rho$ range are indicated with colors corresponding to their respective markers.}
    \label{fig:drho_piKp}
\end{figure}

As detailed in Ref.~\cite{Kincses:2025izu}, the $D(\boldsymbol{\rho})$ distribution clearly shows a core--halo distinction, i.e., a single Lévy-stable distribution can describe the \textit{core} part below $\sim100~{\rm fm}$, while the \textit{halo} part at higher $\boldsymbol{\rho}$ values is defined by the weak decays of a few long-lived resonances. The latter can produce particles even at several cm distances, and affects primarily the correlation strength parameter $\lambda$, as detailed in Section 2.1 of Ref.~\cite{Kincses:2025izu}. This is true for both pion, kaon, and proton pairs; however, the core-halo ratio can be quite different between the particle species. An example comparison of such source distributions (a projection in the \textit{out} direction) is shown in Figure~\ref{fig:drho_piKp} up to 1~m distance. Typical parent particles (resonances) producing protons, kaons, and pions are also shown in the Figure. The decay product protons are either created by very short lived resonances with a few fm mean lifetime (e.g., $N^*$ nucleon resonances, or $\Delta$ particles), or from baryon resonances with extremely long lifetime such as $\Lambda_c,\Sigma,\Lambda$. For kaons, the main resonance sources in the intermediate range between a few~fm-s and 100~fm are the $K^*$ resonances and the $\phi$ meson, while at large distances the main contribution is from the $D$ mesons. For pions, as discussed in Ref.~\cite{Kincses:2025izu}, there is a much larger variety of resonance sources, even in ranges where kaons and protons are not typically produced. 

\begin{figure*}
    \centering
    \includegraphics[trim=0 20bp 0 0, clip, width=0.9\linewidth]{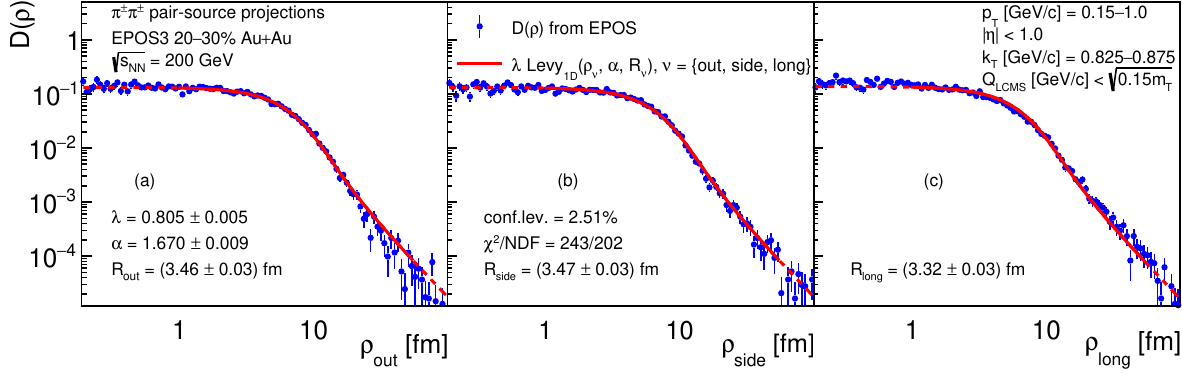}\\[-1ex]
    \vspace{-2.5mm}
    \includegraphics[trim=0 20bp 0 0, clip, width=0.9\linewidth]{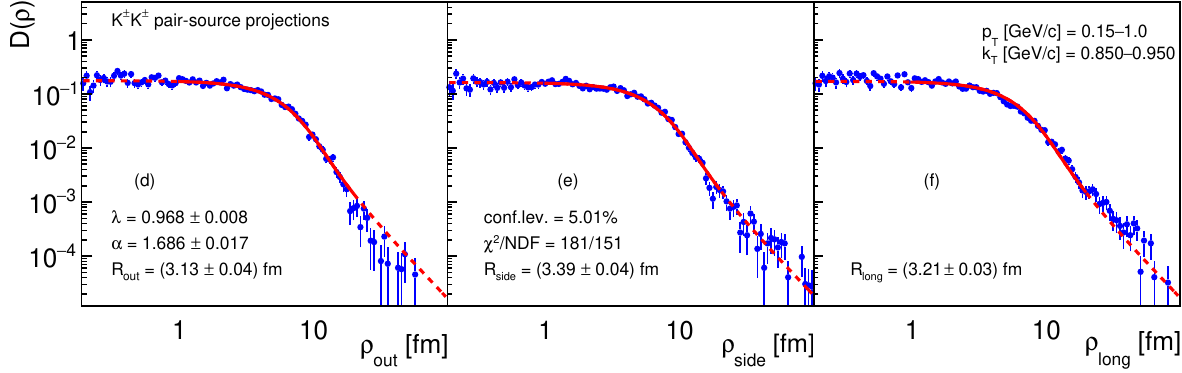}\\[-1ex]
    \vspace{-2.5mm}
    \includegraphics[trim=0 0 0 0, clip, width=0.9\linewidth]{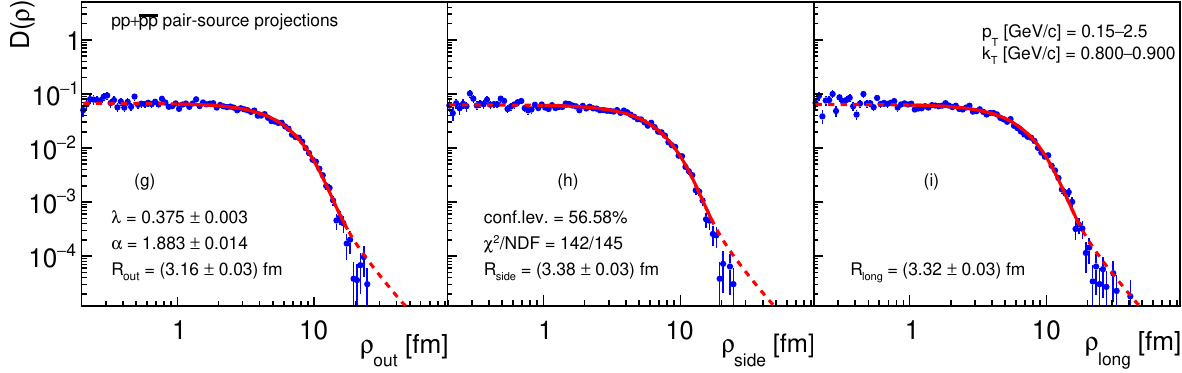}
    
    \caption{Example fits to projected $D(\boldsymbol{\rho})$ distributions of pion (a)-(c), kaon (d)-(f), and proton (g)-(i) pairs, in the $20{-}30\%$ centrality class. Different particle species are shown in different rows, while the projections in the three main directions (\textit{out,side,long}) are shown in different columns. The fit function is shown with a solid red line within the applied fit range, and with a dashed line extrapolated outside of the fit range.}
    \label{fig:examplefit}
\end{figure*}

After the source distributions are constructed from the simulated events for the different $m_T$ and centrality classes, the next step is to fit them with an elliptically contoured three-dimensional Lévy-stable source-shape assumption and obtain the source parameters $\alpha, R_{\rm out,side,long}, \lambda$ as a function of the average transverse mass $m_T$ and centrality. The fitting method is described in detail in Section~2.2 of Ref.~\cite{Kincses:2025izu}. We follow the exact same methodology of simultaneously fitting the three one-dimensional projections of the three-dimensional distribution with a single Lévy-exponent parameter $\alpha$ and strength parameter $\lambda$, but different scale parameters corresponding to the projection direction ($R_{\rm out,side,long}$). An example for such a fit is shown in Figure~\ref{fig:examplefit}.

To estimate the systematic uncertainties arising from variations of different analysis settings, we repeated such fits for a number of different conditions and calculated the difference compared to the default setting. The total systematic uncertainty is calculated as the squared sum of the uncertainties of the various sources, such as the number of events merged ($N_{\rm events}$), fit limits ($\rho_{\rm max}^{\rm fit}$), and pair relative momentum selection ($Q_{\rm LCMS}^{\rm max}$). As mentioned above, in this analysis, we extended the range of investigation in $k_T$ for pion pairs, therefore a higher default value of $N_{\rm events}$ was chosen compared to those described in Section~2.3 of Ref.~\cite{Kincses:2025izu}. The default values are listed below (with the variations in brackets), separately for pion, kaon and proton pairs.
\begin{itemize}
    \item $N_{\rm events}$, pions:
    \begin{itemize}
        \item $0{-}10\%:$ 200 (100, 30\,000)
        \item $10{-}20\%:$ 500 (300, 30\,000)
        \item $20{-}30\%:$ 1\,000 (500, 30\,000)
    \end{itemize}
    \item $N_{\rm events}$, kaons:
    \begin{itemize}
        \item $0{-}10\%:$ 1\,000 (200, 30\,000)
        \item $10{-}20\%:$ 3\,000 (1\,000, 30\,000)
        \item $20{-}30\%:$ 5\,000 (3\,000, 30\,000)
    \end{itemize}
    \item $N_{\rm events}$, protons:
    \begin{itemize}
        \item $0{-}10\%:$ 3\,000 (1\,000, 30\,000)
        \item $10{-}20\%:$ 10\,000 (7\,000, 30\,000)
        \item $20{-}30\%:$ 30\,000 (20\,000, 30\,000)
    \end{itemize}
    \item $Q_{\rm LCMS}^{\rm max}$: $\sqrt{A\cdot m_T}$
    \begin{itemize}
        \item pions: $A = 0.15\,(0.05, 0.25)~[{\rm GeV}]$
        \item kaons: $A = 0.15\,(0.05, 0.25)~[{\rm GeV}]$
        \item protons: $A = 0.15\,(0.10, 0.20)~[{\rm GeV}]$
    \end{itemize}
    \item $\rho_{\rm max}^{\rm fit}$: $\sqrt{B/m_T}$
    \begin{itemize}
        \item pions: $B=2500\,(1600, 3600)~[{\rm fm^2\cdot GeV/}c^2]$
        \item kaons, protons: $B=400\,(225, 625)~[{\rm fm^2\cdot GeV/}c^2]$
    \end{itemize}
\end{itemize}
A summary of the $m_T$ and centrality averaged systematic uncertainties is shown in Table~\ref{tab:syst}. In addition to the settings mentioned above, we also investigated the dependence on the choice of the number of bins used for the $D(\boldsymbol{\rho})$ histograms and found that for pions and kaons it is a negligible ($< 0.1\%$) source of uncertainty, while for proton pairs it is slightly more relevant, but still corresponds to a less than $0.15\%$ change in the parameters on average. 

\begin{table*}[h]
\centering
\caption{Asymmetric relative systematic uncertainties in percentages, averaged over $m_T$ and centrality, shown separately for the three investigated particle species, and three different uncertainty sources for all source parameters.}
\label{tab:syst}
\resizebox{\textwidth}{!}{%
\begin{tabular}{llcccccccccccc}
\toprule
\textbf{Particle} & \textbf{Systematics type} & $\alpha_{\downarrow}$ & $\alpha_{\uparrow}$ & $R_{\text{out},\downarrow}$ & $R_{\text{out},\uparrow}$ & $R_{\text{side},\downarrow}$ & $R_{\text{side},\uparrow}$ & $R_{\text{long},\downarrow}$ & $R_{\text{long},\uparrow}$ & $R_{\text{avg},\downarrow}$ & $R_{\text{avg},\uparrow}$ & $\lambda_{\downarrow}$ & $\lambda_{\uparrow}$ \\ \midrule
\multirow{4}{*}{proton} & fit range & 3.1 & 1.1 & 0.1 & 0.5 & 0.1 & 1.1 & 1.2 & 0.5 & 0.1 & 0.3 & 0.3 & 0.9 \\
                        & $N_{\rm events}$   & 1.0 & 0.3 & 0.2 & 0.3 & 0.1 & 0.4 & 0.3 & 0.2 & 0.1 & 0.2 & 0.4 & 1.1 \\
                        & $Q_{\rm LCMS}$     & 0.8 & 0.7 & 1.7 & 1.2 & 4.4 & 4.0 & 4.5 & 4.2 & 3.5 & 3.2 & 0.8 & 0.6 \\
                        & total     & 3.8 & 1.5 & 1.7 & 1.5 & 4.3 & 4.2 & 4.7 & 4.3 & 3.5 & 3.2 & 1.0 & 1.7 \\ \midrule
\multirow{4}{*}{kaon}   & fit range & 1.7 & 0.6 & 0.4 & 1.1 & $< 0.1$ & 0.4 & 1.4 & 0.6 & 0.1 & 0.1 & 0.1 & 0.5 \\
                        & $N_{\rm events}$   & $< 0.1$ & 0.1 & 0.1 & $< 0.1$ & 0.1 & $< 0.1$ & 0.1 & $< 0.1$ & 0.1 & $< 0.1$ & $< 0.1$ & $< 0.1$ \\
                        & $Q_{\rm LCMS}$     & 1.5 & 1.9 & 2.0 & 1.9 & 5.1 & 4.9 & 6.4 & 5.8 & 4.5 & 4.2 & 0.1 & 0.1 \\
                        & total     & 2.4 & 2.0 & 2.1 & 2.3 & 5.1 & 4.9 & 6.6 & 5.9 & 4.5 & 4.2 & 0.2 & 0.5 \\ \midrule
\multirow{4}{*}{pion}   & fit range & 0.1 & 0.1 & $< 0.1$ & 0.1 & $< 0.1$ & $< 0.1$ & 0.1 & 0.1 & $< 0.1$ & $< 0.1$ & $< 0.1$ & $< 0.1$ \\
                        & $N_{\rm events}$   & 0.1 & $< 0.1$ & 0.1 & $< 0.1$ & $< 0.1$ & $< 0.1$ & $< 0.1$ & $< 0.1$ & $< 0.1$ & $< 0.1$ & $< 0.1$ & 1.1 \\
                        & $Q_{\rm LCMS}$     & 0.6 & 0.7 & 1.0 & 1.0 & 3.0 & 2.7 & 5.1 & 4.7 & 3.0 & 2.8 & 0.1 & 0.1 \\
                        & total     & 0.6 & 0.7 & 1.0 & 1.0 & 3.0 & 2.7 & 5.1 & 4.7 & 3.0 & 2.8 & 0.1 & 1.1 \\ \bottomrule
\end{tabular}%
}
\end{table*}

\section{Results and discussion}\label{s:results}

\begin{figure*}
    \centering
    \includegraphics[width=\linewidth]{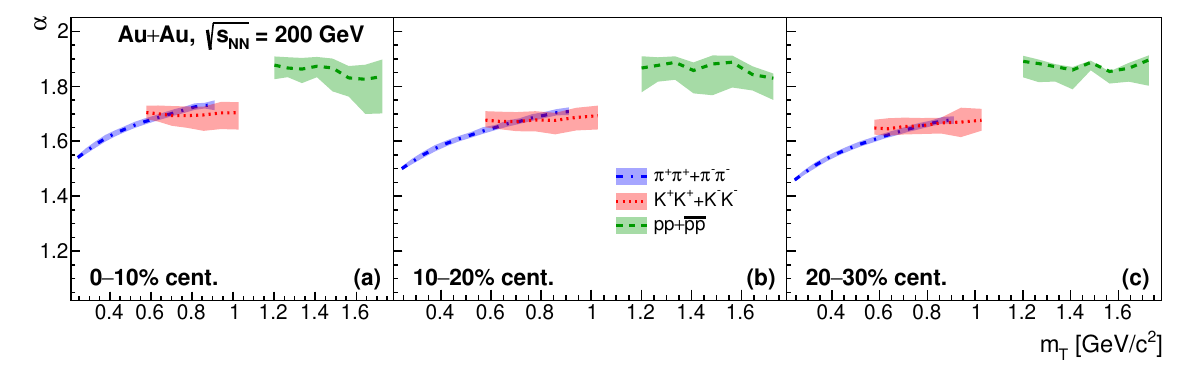}
    \caption{Lévy exponent parameter $\alpha$ as a function of average transverse mass $m_T$. The three centrality classes are shown in separate panels, while on a given panel, the values corresponding to the three different particle species are shown with different colors and line styles. The bands around the dashed lines correspond to the total systematic uncertainty.}
    \label{fig:alpha}
\end{figure*}

The Lévy-exponent parameter $\alpha$ represents the power-law behavior of the source and is mainly affected by various processes that occur throughout the hadronic scattering phase of the collisions. Altogether, these processes can be described within the framework of Lévy walks, as described in Ref.~\cite{Kincses:2024lnv}. This includes elastic and inelastic scatterings, as well as resonance decays and resonance formation (coalescence) processes. The limiting case of stable distributions is $\alpha = 2$, which corresponds to the normal (Gaussian) distribution. Larger values of $\alpha$ can also be mathematically understood, but no longer correspond to a valid probability distribution because the characteristic function ceases to be positive-definite. Values of $\alpha$ between 0 and 2 correspond to distributions with a power-law tail. The average transverse mass dependence of $\alpha$ is shown in Figure~\ref{fig:alpha} on separate panels for the three centrality classes. Interestingly, although the resonance decay contributions for pions and kaons are quite different, the resulting $\alpha$ values are compatible within uncertainties in the overlapping $m_T$ range. This suggests a common origin of the source geometry of pions and kaons. For protons, the $\alpha$ values are much closer to the Gaussian limiting case, as there are hardly any particles decaying to protons in the $10{-}100$~fm range, where the power-law tail usually forms. Interestingly, $\alpha$ exhibits a strong increase with $m_T$ for pion pairs, while it is approximately constant for kaon and proton pairs. This might be attributed to the stronger contribution of resonance decays for low-$m_T$ pions. In Ref.~\cite{Csanad:2007fr}, the kaon exponent was observed to be significantly smaller than that of pions, which can be attributed to the fact that in that analysis only elastic scattering of hadrons was considered. As detailed in Ref.~\cite{Kincses:2024lnv}, in a realistic hadronic scattering scenario, where both elastic and inelastic scatterings, as well as resonance decays and formations are considered, the main process driving the value of the exponent is the Lévy walk. In the intermediate range between $10-100$~fm pair distances (which encloses the tail of the core-core pair source distribution), both pions and kaons have similar resonance contributions (from the $\omega$ and $\phi$ mesons), which can explain their similar $\alpha$ values.

\begin{figure*}
    \centering
    \includegraphics[trim=0 20bp 0 0, clip, width=0.9\linewidth]{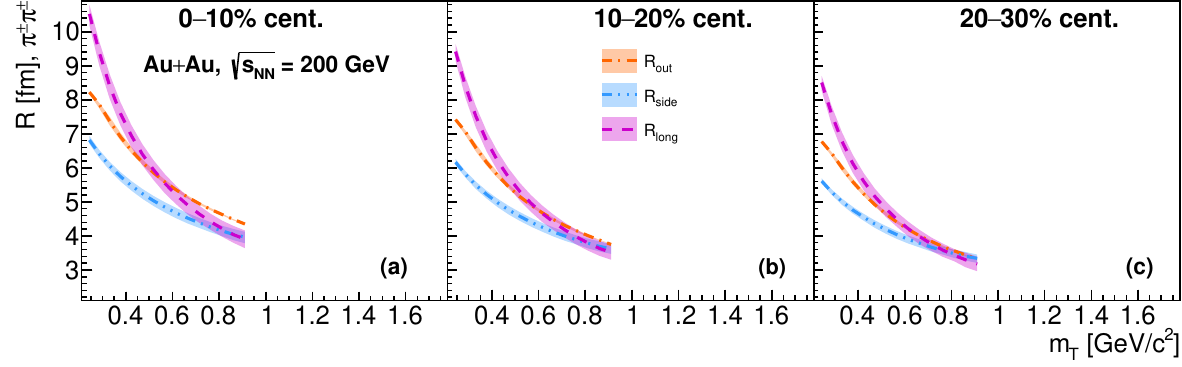}\\[-1ex]
    \vspace{-2.5mm}
    \includegraphics[trim=0 20bp 0 0, clip, width=0.9\linewidth]{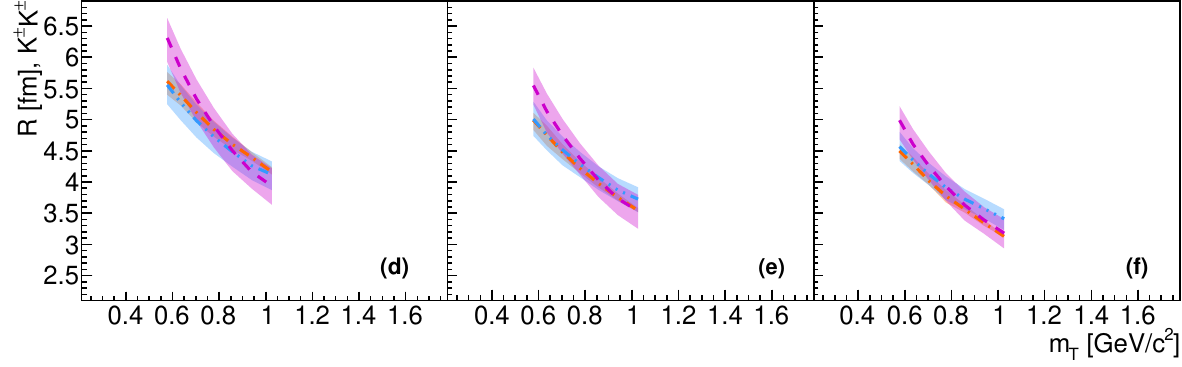}\\[-1ex]
    \vspace{-2.5mm}
    \includegraphics[trim=0 0 0 0, clip, width=0.9\linewidth]{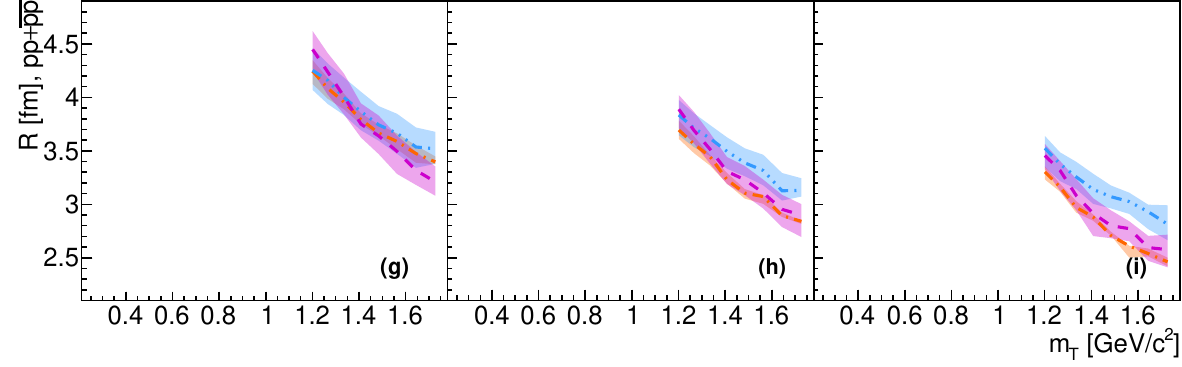}
    
    \caption{Lévy scale parameters $R_{\rm out,side,long}$ as a function of average transverse mass $m_T$. The three centrality classes are shown in separate columns, and the three particle species are shown in separate rows. On a given panel, the values corresponding to the three different directions (\textit{out,side,long}) are shown with different colors and line styles. The bands around the dashed lines correspond to the total systematic uncertainty.}
    \label{fig:R3D}
\end{figure*}

The Lévy-scale parameters $R_{\rm out}$, $R_{\rm side}$, $R_{\rm long}$ are shown in Figure~\ref{fig:R3D}. The particle species are shown in different rows, so that the out-side-long comparisons in a given panel are more clear.  We note in particular that a clear $R_{\rm out}>R_{\rm side}$ ordering is visible for pions, due to prolonged emission duration, which is connected to the Lévy walk phenomenon. This feature weakens for high-momentum particles, i.e., the pion source becomes approximately symmetric in the transverse plane. The kaon and the proton sources exhibit a $R_{\rm out}\approx R_{\rm side}$ relation in general. It is also noteworthy that the side radius becomes larger than the out one for large proton momenta. This may be explained by the stronger space-time ($x-t$) correlations which reduces the out radius, beyond the small increase caused by the emission duration contribution~\cite{Cimerman:2017lmm}. 

\begin{figure*}
    \centering
    \includegraphics[width=\linewidth]{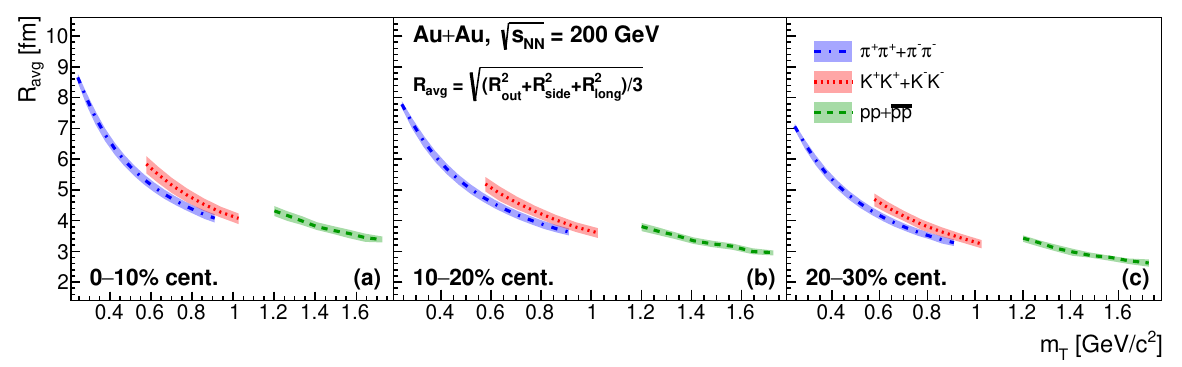}
    \caption{Average Lévy scale parameter $R_{\rm avg}$ as a function of average transverse mass $m_T$. The three centrality classes are shown in separate panels, while on a given panel, the values corresponding to the three different particle species are shown with different colors and line styles. The bands around the dashed lines correspond to the total systematic uncertainty.}
    \label{fig:Ravg}
\end{figure*}

The ${R_{\rm avg} = \sqrt{(R_{\rm out}^2{+}R_{\rm side}^2{+}R_{\rm long}^2)/3}}$ average scale parameter is shown in Figure~\ref{fig:Ravg}. We find that there is no clear $m_T$ scaling between particle species (as suggested by hydrodynamical studies~\cite{Csanad:2008gt}), since the kaon scale parameter is systematically above the pion in the overlapping region. This finding is similar to the experimental measurements of Refs.~\cite{ALICE:2017iga,PHENIX:2015jaj}, where it was found that Gaussian pion and kaon radii scale better with $k_T$ than with $m_T$. Similar trends were also observed in Ref.~\cite{Shapoval:2014wya}, where a hybrid (hydro phase +hadronic rescattering) model was applied. On the other hand, in Ref.~\cite{Kisiel:2014upa} it was investigated what happens to such scalings in a model which includes the hydrodynamic phase as well as statistical hadronization and resonance contribution, but does not include hadronic rescattering. In that case, an approximate $m_T$ scaling of the radii was observed in LCMS. In a subsequent study, it could be investigated whether the $m_T$ scaling violations for Lévy-stable source parameters decrease if hadronic scattering is turned off in EPOS. Interestingly, although all of the mentioned studies utilized the Gaussian approximation, their findings about the scaling behavior among particle species (especially between pion and kaon pairs) are very similar to our observations. This might be due to the fact that the pion and kaon $\alpha$ values are compatible, hence the radii shift compared to the Gaussian approximation would be similar as suggested by Ref.~\cite{Csanad:2024jpy}. Let us also note here that we include three further figures in the Appendix, that provide additional insight into the results. Fig.~\ref{fig:R3D_dirs} shows the scale parameters separately for directions but together for particle species, while Figs.~\ref{fig:R3D_dirs_kt} and~\ref{fig:Ravg_kt} show the scale parameters as a function of $k_T$ instead of $m_T$.

\begin{figure*}
    \centering
    \includegraphics[width=\linewidth]{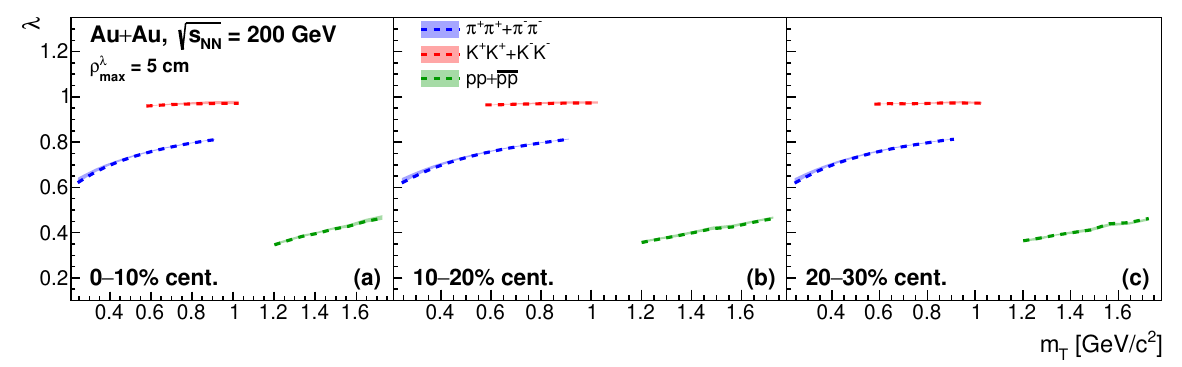}
    \caption{Source strength parameter $\lambda$ as a function of average transverse mass $m_T$. The three centrality classes are shown in separate panels, while on a given panel, the values corresponding to the three different particle species are shown with different colors and line styles. The bands around the dashed lines correspond to the total systematic uncertainty.}
    \label{fig:lambda}
\end{figure*}

The source strength parameter $\lambda$ is shown in Figure~\ref{fig:lambda}. The choice of normalization parameter $\rho^\lambda_{\rm max}$ was 5~cm (see the discussion about it in Section~2.2 of Ref.~\cite{Kincses:2025izu}). A clear ${\lambda_p < \lambda_\pi < \lambda_K}$ ordering can be seen between the different particle species, due to the difference in their respective core-halo ratios, as illustrated also by Fig.~\ref{fig:drho_piKp}. 

\section{Summary and outlook}\label{s:summary}
In this paper, we presented a detailed three-dimensional investigation of the particle species dependence of the pair-distance distribution in Monte-Carlo simulations of Au+Au collisions at 200 GeV per nucleon collision energy, using the EPOS model. We extracted the Lévy source parameters $\alpha$, $R_{\rm out,side,long}$ and $\lambda$ as a function of the average transverse mass $m_T$ and centrality, using the methodology of Ref.~\cite{Kincses:2025izu}. We found that the Lévy exponent $\alpha$ of identical charged pion and kaon pairs is compatible within uncertainties in the overlapping $m_T$ region and exhibits clear non-Gaussian ($\alpha <2$) values across all investigated $m_T$ and centrality. For proton pairs, it has higher values, closer to the Gaussian limit. The Lévy scale parameters do not show a clear scaling $m_T$ (contrary to hydrodynamic predictions), as the kaon parameters are systematically above the pion in the overlapping region. This is consistent with the experimental findings of Refs.~\cite{ALICE:2017iga,PHENIX:2015jaj}, despite the Gaussian approximation used in those analyses. The strength parameter shows a clear ${\lambda_p < \lambda_\pi < \lambda_K}$ ordering, due to the differences of long-lived resonance contribution. These results will provide the basis for a future comparison with experimental data, as Lévy-stable measurements for kaon and proton pairs are currently underway at multiple different experiments. 

Let us also note that as a further continuation of these analyses, we plan to study the source shape of non-identical (pion-kaon and pion-proton) pairs, where the spatial separation distribution is expected to be a convolution of two stable-distributions with different exponent parameters. Such investigations can shed light on interesting emission asymmetries between particle species~\cite{ALICE:2020mkb,ALICE:2025aur,Kisiel:2018wie}.

\begin{appendices}
\section{}\label{secA1}
\begin{figure*}
    \centering
    \includegraphics[trim=0 20bp 0 0, clip, width=0.9\linewidth]{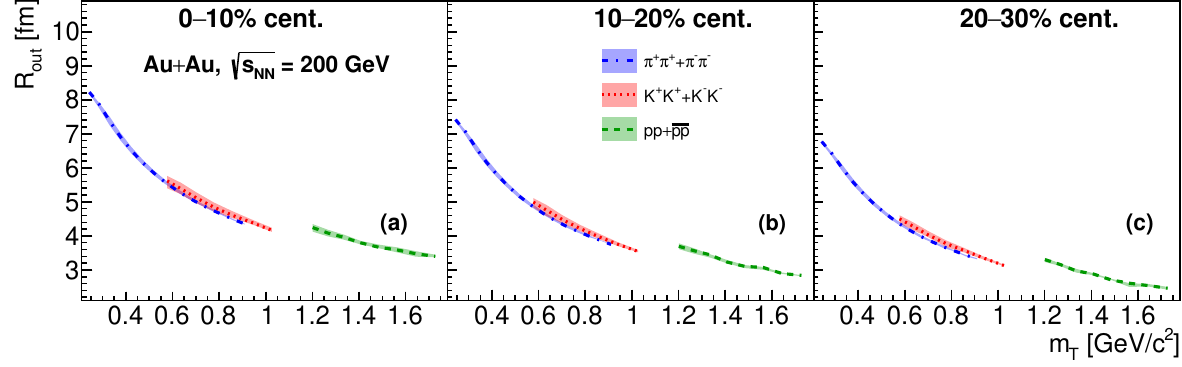}\\[-1ex]
    \vspace{-2.5mm}
    \includegraphics[trim=0 20bp 0 0, clip, width=0.9\linewidth]{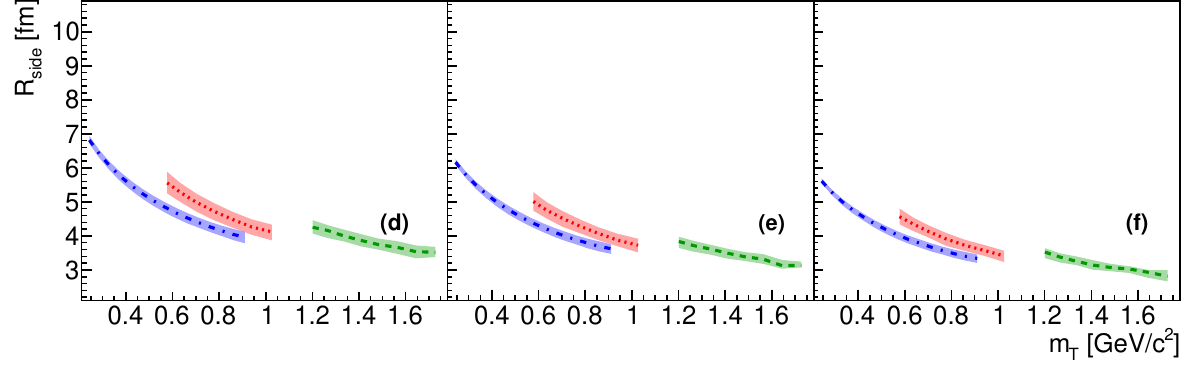}\\[-1ex]
    \vspace{-2.5mm}
    \includegraphics[trim=0 0 0 0, clip, width=0.9\linewidth]{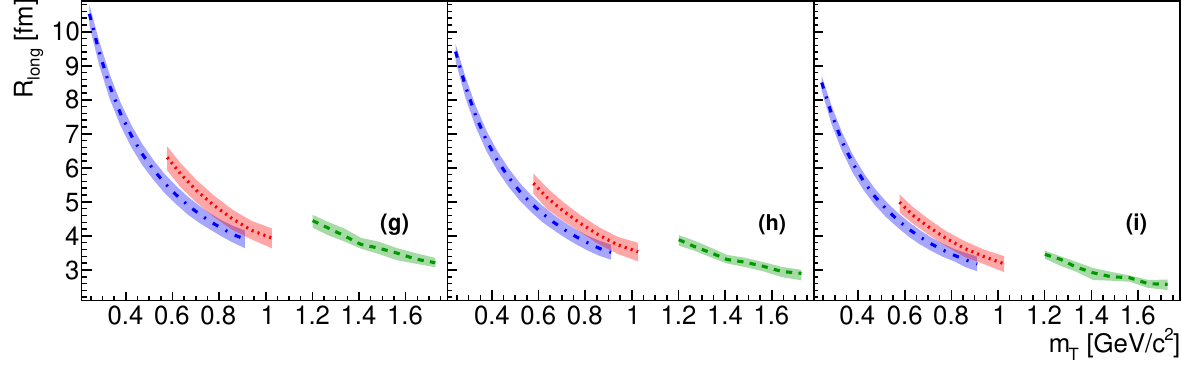}
    
    \caption{Lévy scale parameters $R_{\rm out}$, $R_{\rm side}$, and $R_{\rm long}$ versus $m_T$, shown separately (in panels (a)-(c), (d)-(f), and (g)-(i), respectively). Pion, kaon, and proton parameters are shown in blue, red, and green, respectively. Each result is shown for three centralities, 0--10\%, 10--20\%, and 20--30\%, in the three consecutive columns.}
    \label{fig:R3D_dirs}
\end{figure*}

\begin{figure*}
    \centering
    \includegraphics[trim=0 20bp 0 0, clip, width=0.9\linewidth]{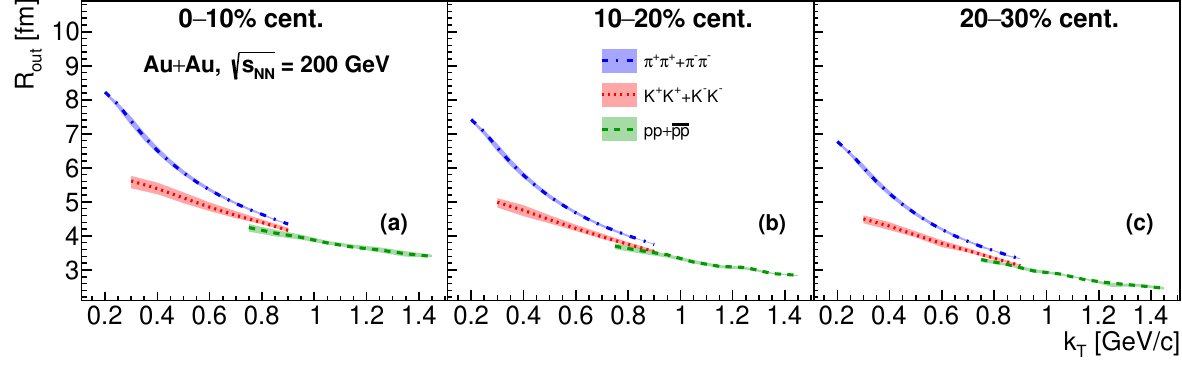}\\[-1ex]
    \vspace{-2.5mm}
    \includegraphics[trim=0 20bp 0 0, clip, width=0.9\linewidth]{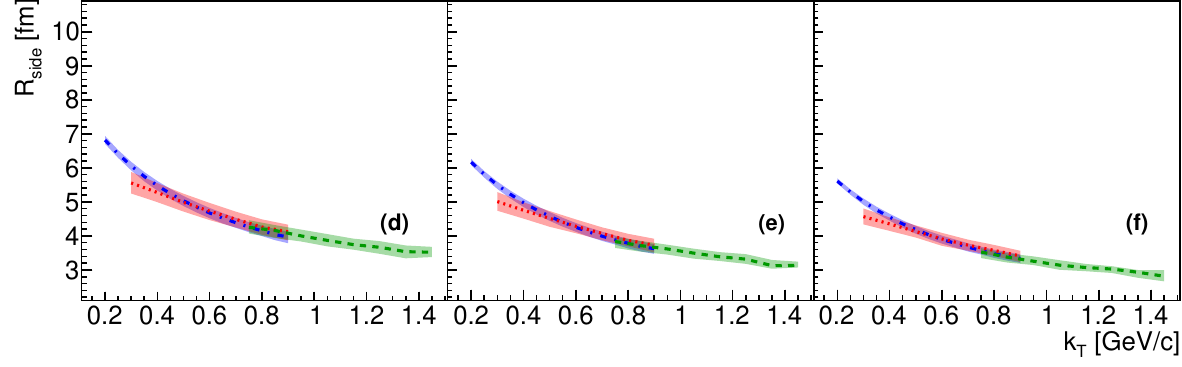}\\[-1ex]
    \vspace{-2.5mm}
    \includegraphics[trim=0 0 0 0, clip, width=0.9\linewidth]{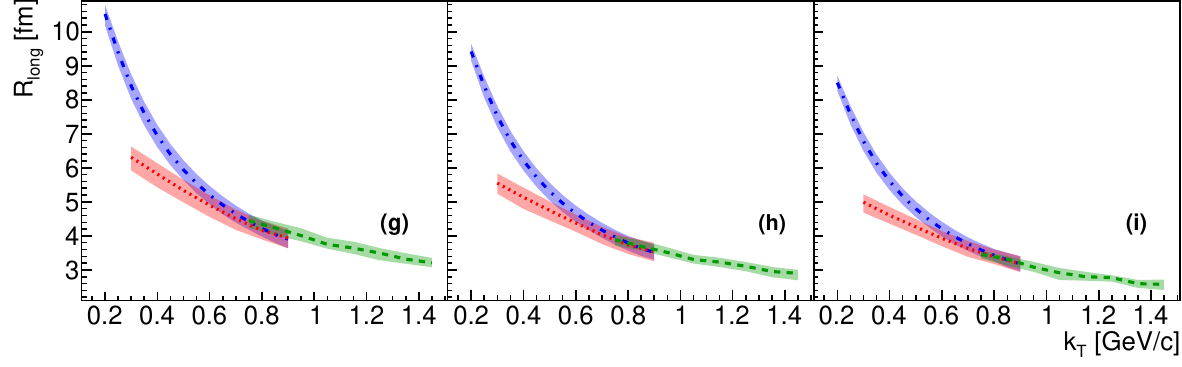}
    
    \caption{Lévy scale parameters $R_{\rm out}$, $R_{\rm side}$, and $R_{\rm long}$ versus $k_T$, shown separately (in panels (a)-(c), (d)-(f), and (g)-(i), respectively). Pion, kaon, and proton parameters are shown in blue, red, and green, respectively. Each result is shown for three centralities, 0--10\%, 10--20\%, and 20--30\%, in the three consecutive columns.}
    \label{fig:R3D_dirs_kt}
\end{figure*}

\begin{figure*}
    \centering
    \includegraphics[width=\linewidth]{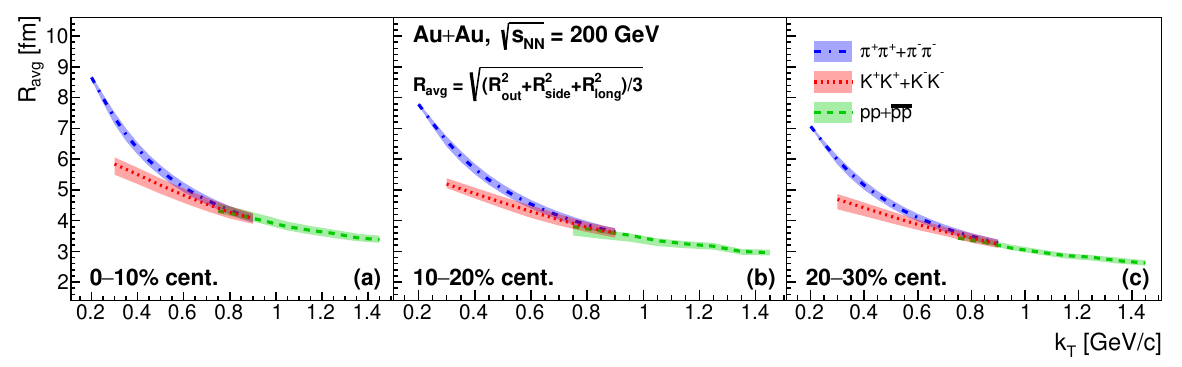}
    \caption{Average Lévy scale parameter $R_{\rm avg}$, shown versus $k_T$ for three centralities, 0--10\%, 10--20\%, and 20--30\%, in panels (a), (b), and (c), respectively. Pion, kaon, and proton parameters are shown in blue, red, and green, respectively.}
    \label{fig:Ravg_kt}
\end{figure*}

In this appendix we include additional figures that are out of the scope of the main discussion, but provide additional insight into the results. Fig.~\ref{fig:R3D_dirs} indicates the Lévy scale parameters $R_{\rm out,side,long}$ separately. It is apparent that $m_T$ scaling is the least violated for $R_{\rm out}$, while the other two directions exhibit a comparable difference between scales for different particle species. Figure~\ref{fig:R3D_dirs_kt} shows the same scale parameters versus $k_T$. It is apparent that while proton and kaon radii match at the same $k_T$, there is a sizeable difference between pion and kaon results, especially for the out and long directions, where $m_T$ scaling appears to be much less violated. The same conclusion can be drawn from Fig.~\ref{fig:Ravg_kt}, showing the $k_T$ dependence of the average Lévy scale $R_{\rm avg}$. 

\end{appendices}

\bmhead{Acknowledgements}

This research was funded by the NKFIH grants PD-146589, 2026-4.1.1-MISSZIÓK-2026-00005, and NKKP ADVANCED 152097. L. K. was supported by the EKÖP-26 University Research Scholarship Program of the Ministry of Science and Technology from the source of the National Research, Development and Innovation Fund.


\bibliography{sn-bibliography}

\end{document}